# High-temperature spin-wave propagation in BiFeO$_3$: relation to the Polomska transition

**Ashok Kumar[1], J. F. Scott[1,2], R. S. Katiyar[1]**

[1]Department of Physics and Institute for Functional Nanomaterials, University of Puerto Rico, San Juan, Puerto Rico, USA, PR-00936-8377

[2]Department of Physics, Cavendish Laboratory, Cambridge University, Cambridge CB3 OHE, United Kingdom

**Abstract:**

In bismuth ferrite thin films the cycloidal spiral spin structure is suppressed, and as a result the spin-wave magnon branches of long wavelength are reduced from a dozen to one, at ω = 19.2 cm$^{-1}$ (T=4K). This spin wave has not been measured previously above room temperature, but in the present work we show via Raman spectroscopy that it is an underdamped propagating wave until 455 K. This has important room-temperature device implications. The data show that ω(T) follows an S=5/2 Brillouin function and hence its Fe$^{+3}$ ions are in the high-spin 5/2 state and not the low-spin S=1/2 state. The spin wave cannot be measured as a propagating wave above 455 K. We also suggest that since this temperature is coincident with the mysterious "Polomska transition" (M. Polomska et al., *Phys. Stat. Sol. A* 23, 567, (1974)) at 458+/-5 K, that this may be due to overdamping.

**Corresponding Author:** E-mail: *J F Scott jfs32@hermes.cam.ac.uk, R S katiyar, rkatiyar@uprrp.edu

The production of high-quality thin films of bismuth ferrite has opened up the possibility of making room-temperature devices that rely upon spin wave propagation. In 2008 the study of spin waves via magnon Raman spectroscopy began [1-5], and more recently a review was published on its device applications [6]. Foremost among the potential device applications are THz emitters [7,8,9], and for the majority of devices (including electric-field tunable magnetic devices) the existence of room-temperature spin waves (magnons) that are propagating modes (as opposed to overdamped or diffusive modes) is important [10,11,12]. Here we report propagating spin waves in $BiFeO_3$ up to 450 K. Although in bulk there are numerous magnon branches at long wavelength that scatter light in the Raman effect (due to the cycloidal spin structure), in thin films this cycloid is suppressed and there is only a single magnon branch in the Raman spectra. Loudon and Fleury [13,14] provided the basic theory of magnon spectroscopy and damping. In general, systems with Fe or Co ions can exhibit unquenched orbital angular momentum; but despite the fact that they therefore are not pure spin systems, their temperature dependence and damping are not very different from those in pure spin systems such as Mn compounds [15]. Low-energy spin waves are observed in several ferro/antiferromagnetic systems and well explained with a phenomenological theory [16,17,18].

Here we report the temperature dependent softening of spin waves, their suppression in thin-film form, correlation with a magnetic transition temperature, and dielectric loss. Softening of spin waves follows a modified Brillouin function with S=5/2 (which is mean field with critical susceptibility exponent β=1/2).

Pre-patterned platinum interdigital electrodes were procured from NASA Glenn Research Center's electronics division having dimension 1900 μm (length), 15 μm (interdigital spacing) and 150 +/- 25 nm (height) with 45 parallel capacitors in series on sapphire substrates. Fig. 1 (a, b, c) shows the cartoon of spin waves, their suppression to a spin arrangement similar to that in conventional two-sublattice antiferromagnetic systems, and the in-plane view of the BFO thin films on the pre-patterned interdigital electrodes. The details of crystal phase formations, surface morphology, impurities and electrical testing equipments and parameters were presented in previous report [1,5]. BFO thin films were grown utilizing an excimer laser (KrF, 248 nm) with a laser energy density of 1.5 J/cm$^2$, laser rep-rate of 10 Hz, substrate temperature 650°C and oxygen pressure at 80 mTorr. Micro-Raman spectra were recorded in the backscattering geometry using 514.5 nm monochromatic radiations over wide range of temperature utilizing low temperature cryostage from Linkam.

Figure 2 demonstrates the variation of low-energy spin waves at various temperatures. Only one spin wave was observed in the low-frequency Raman spectra which is sharp and can be clearly seen beyond the baseline noise background. These low-energy spin waves are underdamped until 450 K with a sharp magnon peak. Spin waves shift towards lower frequencies with increase in temperature; our due to stray light and spectrometer characteristics, our experimental limitation was to observe the Raman spectra only above 10 cm$^{-1}$.

Fig. 3 shows the scaling behaviour magnon frequencies ω(T) versus temperature in one of our films. Note that the frequency data ω(T) satisfy a mean-field S=5/2 high-spin Brillouin function better than a low-spin S=1/2 Brillouin function. It is already known from a series of elegant papers by Gavriliuk et al. in Moscow [19-22] that the $Fe^{+3}$ ions in the insulating state of bismuth ferrite have S=5/2, whereas the metallic state has S=1/2, so this only confirms previously established conclusions. However, this is the first time such conclusions can be inferred purely from Raman magnon spectroscopy in any material, so that it provides a nice pedagogical example.

BFO thin films show only single spin wave compare to the electromagnons (several spin excitations) in single crystal, compel us to explain the magnon frequency to a single quantum excitation, a molecular field theory can provide an interpretation based on the macroscopic parameters, $H_A$ and $H_E = |\lambda M_1| = |\lambda M_2|$, where $H_A$ and $H_E$ indicates anisotropy and exchange field[16].

The solution of Bloch equations provide the low lying magnon frequency

$$\omega_m = \gamma(H^2_A + 2H_A H_E)^\beta \quad \text{...............................(1)}$$

$$\gamma = ge/2mc, S = 5/2, g = 2$$

Where γ is the gyromagnetic ratio of $Fe^{3+}$ ions, m is electronic mass, g-factor, c is speed of light, e is the electronic charge.

In the framework of molecular field theory the net magnetization is the superposition of two interpenetrating magnetization $M_1$ and $M_2$ preferentially parallel and antiparallel to $H_A$. It is assumed that the $H_A$ and $H_E$ is the proportional to the saturation values of $M_1$ and $M_2$ given by;

$$M_s = (2g\mu_B S/a^3) B_s(y) \quad \text{...........................................(2)}$$

where $\mu_B$ is the Bohr magneton, a is the lattice constant, $k_B$ is Boltzmann constant, and $B_S(y)$ the Brillouin function.

$$B_J = \frac{2S+1}{2S}ctnh\left(\frac{(2S+1)y}{2S}\right) - \frac{1}{2S}ctnh\left(\frac{y}{S}\right)$$

for $y = \frac{\mu_B}{k_B T} <<< 1$  $\coth y = \frac{1}{y} + \frac{y}{3} - \frac{y^3}{45}$

$$y = \frac{g\mu_B S}{\kappa_B T}(H_A + H_E) \equiv \frac{g\mu_B S}{\kappa_B T}(\lambda+\mu)M_S \quad \text{...............(3)}$$

$$k_B T_N = (2/3a^3)g^2\mu_B^2 S(S+1)(\lambda+\mu) \quad \text{...............(4)}$$

Solving equation eq. 2 by assuming that $\omega_M$ is proportional to the $M_s$ and equal to 19.2 cm$^{-1}$ at T = 0 K and $T_N$ ~ 631.5 K (power law fitting of experimental data). We have utilized the equation 3 and 4 to solve the equation 2 and Brillouin function. This theoretical model has been used to see the softening of low lying magnon for most of the antiferomagnetic systems. Our experimental data fitted well with the theory very near to spin S=5/2. Darby has obtained the numerical values and the equations for the spontaneous magnetization and their softening behaviour of for magnetic system with different spin behaviour [17].

Near the Neel temperature $T/T_N \rightarrow 1(y << 1)$ Brillouin function and scaled magnetization will follow the equation

$$B_S(y) = \frac{(2S+1)^2 - 1}{(2S)^2}\frac{y}{3} - \frac{(2S+1)^4 - 1}{2S^4}\frac{y^3}{45} + \text{...............(5)}$$

$$\left(\frac{M}{M_0}\right)^2 = \frac{10}{3}\frac{(S+1)^2}{(S+1)^2 + S^2}\left(1 - \frac{T}{T_N}\right) + \text{...............(6)}$$

whereas near the absolute 0 K temperature $T/T_N \to 0 (y \gg 1)$ Brillouin function and scaled magnetization can be obtained from the equation 7 and 8.

$$B_S(y) = 1 - \frac{1}{S}\exp\left(-\frac{y}{S}\right) + \ldots \ldots (7)$$

$$\left(\frac{M}{M_0}\right) = 1 - \frac{1}{S}\exp\left(-\frac{3}{S+1}\frac{T_N}{T}\right) + \ldots \ldots (8)$$

Equation 6 and 8 together offer the different values of magnetization and Brillouin function near low temperature (absolute zero Kelvin) and Neel point respectively. Utilizing both equation one can sketch the complete range of magnetization verses temperature scaling behaviour.

Theoretical modelling of magnetization for different spin behaviours can be obtained from equation 6 and 8.

It is worth mentioning that magnon frequency is directly proportional to magnetization with a factor $\gamma = ge/2mc$ that also involves the spin factor.

$$\omega_m = \gamma M^\beta \quad \gamma = ge/2mc$$

$S = 5/2, g = 2$ for high spin state

But for the low spin the value of 1.5<g <2, we consider g=1.5 in our fitting parameters

$$g_J = \frac{3}{2} + \frac{S(S+1) - L(L+1)}{2J(J+1)}$$

We have utilized the low temperature spin-wave theory since experimental data only reached up to the 70% of the Neel temperature. The scaling behaviour well follow the S=5/2 for both low temperature spin wave theory and modified molecular field theory near the Neel temperature. S=1/2 is very unlikely to fit the experimental data in both cases. It is worth mentioning that our spin wave softening matches very well with the recent observation of the magnon softening probed by terahertz spectroscopic studies [9].

The magnon damping is of special interest (Fig.3): As noted above, it is modest until ca. 150 K above ambient. We find that it is no longer measurable as a propagating wave above 450+/-10 K. It is not unusual for magnons to become overdamped at temperatures of order 70-80% T(Neel), but in the case of bismuth ferrite, this might not be entirely coincidental, since a yet enigmatic phase transition has been reported [23,24] in some specimens at 458+/-5 K. If this is indeed the temperatures of a subtle structural phase transformation (surface [25] or bulk), then one might expect anomalous damping at nearby temperatures. In order to test this coincidence, we present in the final section a complementary study of in-plane dielectric loss, which confirms the strong connection to the Polomska transition.

Temperature dependent line widths of the magnons are presented in the Fig. 4. These show an anomalous decrease in the spin linewidth near the so-called spin reorientation transition (210 K) but with non-monotonic behaviour, increasing at higher temperatures. The detailed phenomena near the presumed magnetric transition at ca. 210K are discussed in earlier reports [1-5]. It is interesting that the in-plane loss tangent of BFO is very small up to 450 K (only 1-2% dielectric loss was observed as can be seen from Fig.5). A reproducible broad loss peak ("hump") was observed near the 210K magnetic transition for wide range of frequency. The detailed in-plane dielectric properties of BFO will be presented elsewhere. It is not a coincidence that the dielectric loss also starts increasing above 450 K where magnon disappears as a propagating mode.

In summary, we report for the first time high temperature spin wave propagation and temperature dependence in BFO thin films. The spin waves remain underdamped until 450 K with a sharp magnon peak. Propagating spin waves were not observed at elevated temperatures. The magnon frequency follows an S=5/2 Brillouin function and hence its $Fe^{+3}$ ions are in the high-spin 5/2 state. The linewidth demonstrates anomalous decrease near the magnetic transition temperature but increases strongly at higher temperatures. A dielectric peak loss broad in temperature was observed near the onset of the 210 K magnetic transition over wide temperature range, and not coincidentally, it starts increasing rapidly above the 450 K Polomska transition, at which the spin waves no longer are observed as propagating waves.

**Acknowledgement:**

This work was supported in parts by DOE-DE-FG02-08ER46526 and IFN-NSF-RII 07-01-25grants.

**Figure captions:**

Fig.1 Schematic diagram and cartoon of the spin waves, spin wave generation by anti ferromagnetic materials, and the in-plane view of the BFO thin films. (a) Spin wave propagation in the BFO single crystal, (b) suppression of spin waves with several frequencies to a single frequency and the behavior is similar to the magnon propagation in antiferromagnetic materials, an incident photon interact with an electron having spin s and force it to hop with the neighboring site leaving a hole and creating a double occupancy in the intermediate state with energy (E). Particle of the double state with opposite spin hops back to the hole site liberating a photon with different energy leaving behind a locally disturbed antiferromagnet. (c) schematic diagram of the in plane view of the BFO thin films grown on the pre patterned interdigited electrode.

(2) Low lying spin waves probed by Raman spectroscopy at various temperatures (80 K to 450 K), Magnon frequency shifted towards lower frequency side with increase in temperature.

(3) Magnon frequency versus temperature in bismuth ferrite thin films. Experimental values of spin wave frequency are presented by solid dots with the experimental error bars line. The curves are least squares fits to S=1/2 (low-spin) (top solid line) and S=5/2 (high-spin) for the magnon frequency for the $Fe^{+3}$ ions utilizing low temperature spin theory (equation 1 & 8). The theoretical fitting of the complete scaling behavior of the experimental data is shown by the last curve (green solid line), that fitted well utilizing modified mean molecular field model. Note that magnon frequency theoretically differs somewhat from magnetization M(T).

(4) Temperature dependent linewidths of magnon frequency showing decrease in linewidth near the 210K magnetic transition temperature and higher temperature increase.

(5) In-plane loss tangent (at 10 kHz) of BFO thin films showing broad maximum near the onset of magnetic phase transition temperature and sharp increase above the polomska transition near 458K.

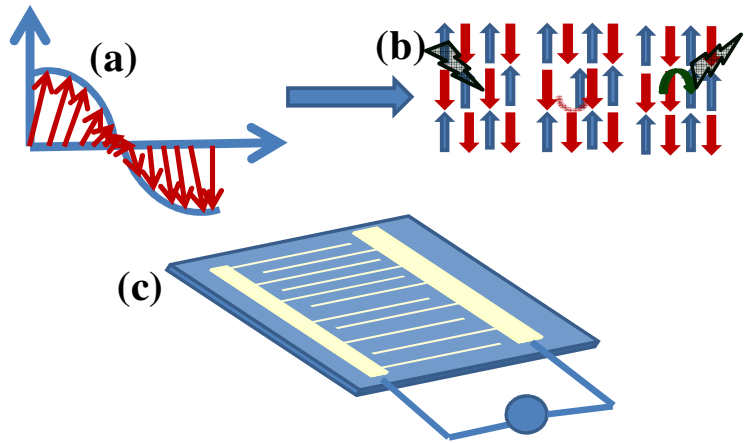

**Fig.1**

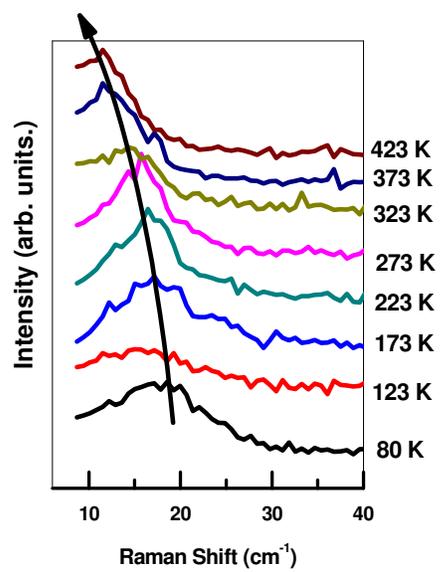

**Fig.2**

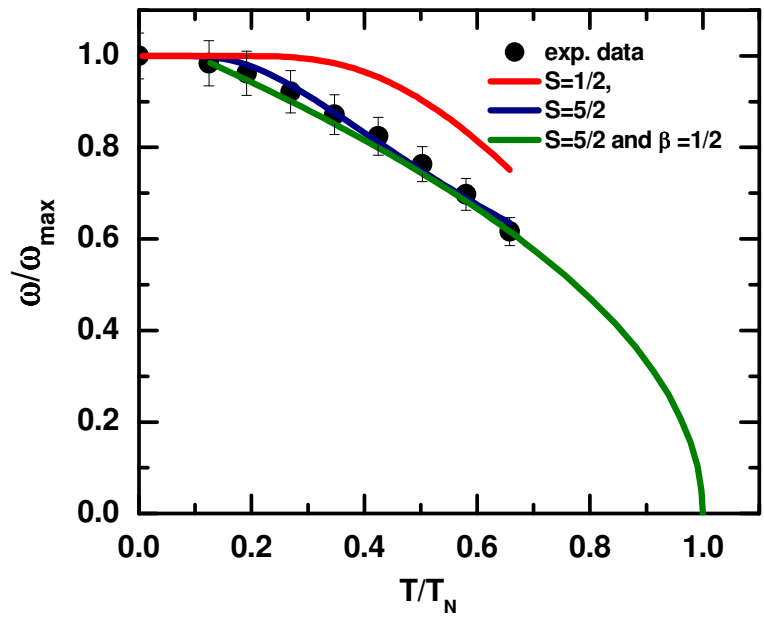

**Fig.3**

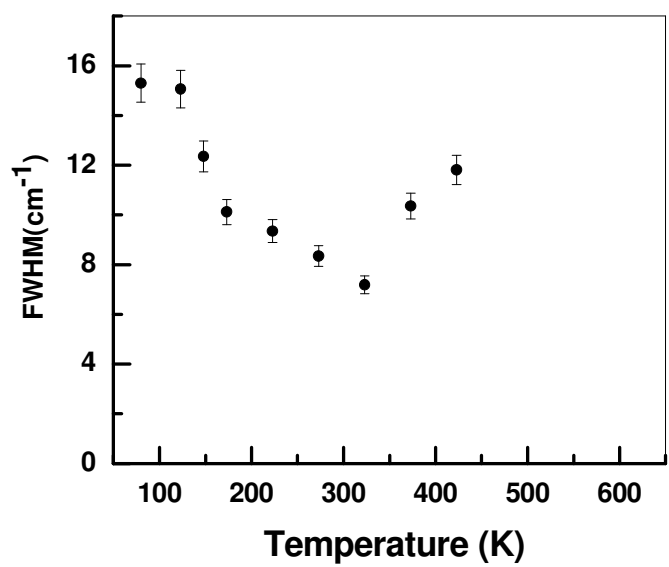

**Fig.4**

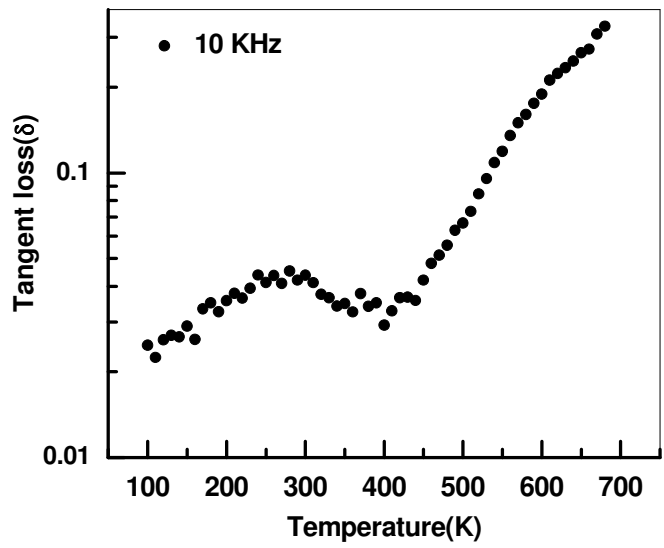

**Fig.5**